\documentclass[journal]{IEEEtran}
\ifCLASSINFOpdf
  \usepackage[pdftex]{graphicx}
\else
   \usepackage[dvips]{graphicx}
\fi
\usepackage[cmex10]{amsmath}
\usepackage{array}
\usepackage{graphicx}
\usepackage{epstopdf}
\usepackage{xargs}
\usepackage{ifthen}
\usepackage{extpfeil}
\usepackage{tikz}
\usepackage{dsfont}
\usepackage{pgf}
\usepackage{stmaryrd}
\usepackage[verbose]{wrapfig}
\usepackage{xcolor}
\usepackage{pbox}
\usepackage{url}
\usetikzlibrary{arrows}
\usepackage{amssymb}
\usepackage{amsmath}
\usepackage{soul}
\soulregister\cite7
\soulregister\ref7
\soulregister\pageref7
\usepackage{verbatim}
\usepackage{hyperref}
 \usepackage[percent]{overpic}
\usepackage{algorithm}
\usepackage[noend]{algpseudocode}
\usepackage[T1]{fontenc}   % T1 font encoding that allows for proper display of some symbols, incl. '<' and '>'
\usepackage[noadjust]{cite} % use this if the citations in the body of the paper have errors like [26,?,?] ? e.g., needed to be used when submitting to a journal and citing wasn't working
\usepackage{authblk}
\usepackage{pdflscape}
\usepackage{geometry}
\usepackage{caption}
\usepackage{breqn}
\usepackage{setspace}
\usepackage[bottom]{footmisc}

\usepackage[absolute]{textpos}
\usepackage{atbegshi}

\hypersetup{
    colorlinks=True,       % false: boxed links; true: colored links
    urlcolor=blue           % color of external links
}

\renewcommand{\figurename}{Fig.}

\renewcommand{\hl}[1]{#1}
\newcommand{\hlc}[1]{#1}
\begin{document}
\title{Solving the Optimal Trading Trajectory Problem Using a Quantum Annealer}

\author{Gili Rosenberg,
        Poya Haghnegahdar,
        Phil Goddard\textsuperscript{*},
        Peter Carr,
        Kesheng Wu,
        Marcos~L\'{o}pez~de~Prado% <-this % stops a space
\thanks{Copyright (c) 2016 IEEE. Personal use of this material is permitted. However, permission to use this material for any other purposes must be obtained from the IEEE by sending a request to pubs-permissions@ieee.org.}%this % stops a space
\thanks{G. Rosenberg and P. Goddard\textsuperscript{*} are with 1QBit, \hl{Suite 458--550 Burrard Street, Vancouver, BC V6C 2B5, Canada.}}% <-this % stops a space
\thanks{\hl{P. Haghnegahdar is with the University of British Columbia, 2329 West Mall, Vancouver, BC V6T 1Z4, Canada.}}% , this % stops a space
\thanks{P. Carr is with New York University (NYU) -- Courant Institute of Mathematical Sciences, 251 Mercer Street, New York, NY 10012, USA.}% <-this % stops a space
\thanks{K. Wu is with Lawrence Berkeley National Laboratory, One Cyclotron Road, Berkeley, CA 94720, USA.}% <-this % stops a space
\thanks{M. L\'{o}pez de Prado is with Guggenheim Partners LLC, 330 Madison Avenue, New York, NY 10017, USA and is a Research Fellow at Lawrence Berkeley National Laboratory's Computational Research Division, One Cyclotron Road, Berkeley, CA 94720, USA.}% <-this % stops a space
\thanks{\textsuperscript{*} Corresponding author \hl{(phil.goddard@1qbit.com)}}% <-this % stops a space
}


\markboth{Special Issue on Financial Signal Processing and Machine Learning for Electronic Trading} % ,~Vol.~13, No.~9, September~2015}%
{Shell \MakeLowercase{\textit{et al.}}: Bare Demo of IEEEtran.cls for Journals}



\maketitle

\begin{abstract}
We solve a multi-period portfolio optimization problem using D-Wave Systems' quantum annealer.
We derive a formulation of the problem, discuss several possible integer encoding schemes, and present numerical examples that show high success rates.  The formulation incorporates transaction costs (including permanent and temporary market impact), and, significantly, the solution does not require the inversion of a covariance matrix. The discrete multi-period portfolio optimization problem we solve is significantly harder than the continuous variable problem. 
We present insight into how results may be improved using suitable software enhancements and why current quantum annealing technology limits the size of problem that can be successfully solved today. The formulation presented is specifically designed to be scalable, with the expectation that as quantum annealing technology improves, larger problems will be solvable using the same techniques. \end{abstract}


% \fi
\IEEEpeerreviewmaketitle

%\doublespacing 
\newpage  % Forces intro to be at top of 2nd column in 2-column format

\AtBeginShipout{
\begin{textblock}{12}(2,0.1)
\centering
\scriptsize This is the author's version of an article that has been published in this journal. Changes were made to this version by the publisher prior to publication. \\
The final version of record is available at \url{http://dx.doi.org/10.1109/JSTSP.2016.2574703}
\end{textblock}
\begin{textblock}{14}(1.3,15.8)
\centering
\scriptsize Copyright (c) 2016 IEEE. Personal use is permitted. For any other purposes, permission must be obtained from the IEEE by emailing pubs-permissions@ieee.org.
\end{textblock}
}

\section{The problem}
%\IEEEPARstart{T}{his} demo file is intended to serve as a ``starter file''
%for IEEE journal papers produced under \LaTeX\ using
%IEEEtran.cls version 1.8a and later.
%I wish you the best of success.

 
%\hfill September 17, 2014

\subsection{Introduction}
\IEEEPARstart{C}{onsider} an asset manager wishing to invest $K$ dollars in a set of $N$ assets with an investment horizon divided into $T$ time steps. Given a forecast of future returns and the risk of each asset at each time step, the asset manager must decide how much to invest in each asset at each time step, while taking into account transaction costs, including permanent and temporary market impact costs. 

One approach to this problem is to compute the portfolio that maximizes the expected return subject to a level of risk at each time step. This results in a series of ``statically optimal'' portfolios. However, there is a cost to rebalancing from a portfolio that is locally optimal at $t$ to a portfolio that is locally optimal at $t+1$. This means that it is highly likely that there will be a different series (or, a ``trajectory'') of portfolios that will be ``globally optimal'' in the sense that its risk-adjusted returns will be jointly greater than the combined risk-adjusted returns from the series of ``statically optimal'' portfolios.

Mean-variance portfolio optimization problems are traditionally solved as continuous-variable problems. However, for assets that can only be traded in large lots, or for asset managers who are constrained to trading large blocks of assets, solving the continuous problem yields an approximation, and a discrete solution is expected to give better results. For example, institutional investors are often limited to trading ``even'' lots (due to a premium on ``odd'' lots), \hl{that is, lots that are an integer multiple of a standard lot size, i}n which case the problem becomes inherently more discrete as the trade size increases versus the lot size.  This could occur, for example, due to the trading of illiquid assets. Two common examples of block trading are ETF-creation and ETF-redemption baskets, which can only be traded in large multiples, such as fund units of 100,000 shares each.

The discrete problem is non-convex due to the fragmented nature of the domain, and is therefore much harder to solve than a similar continuous problem. Furthermore, our formulation allows the covariance matrix to be ill-conditioned or degenerate.
This complicates the finding of a solution using traditional optimizers since a continuous relaxation would still be non-convex, and therefore difficult to solve.
%\IEEEpubidadjcol

%\subsubsection{Subsubsection Heading Here}
%Subsubsection text here.

\subsection{Previous work}
\label{sec:previous_work}

The single-period discrete portfolio optimization problem has been shown to be NP-complete, regardless of the risk measure used \cite{kellerer2000selecting, mansini1999heuristic}. Jobst \textit{et al.} \cite{jobst2001computational} showed that the efficient frontier of the discrete problem is discontinuous and investigated heuristic methods of speeding up an exact branch-and-bound algorithm for finding it. \hl{Vielma \textit{et al.} presented a branch-and-bound algorithm and results for up to 200 assets \cite{vielma2008lifted}}. Heuristic approaches, including an evolutionary algorithm, were investigated by other authors \cite{corazza2007existence, kellerer2000selecting, streichert2004evolutionary}. 

Bonami and Lejeune \cite{bonami2009exact} solved a single-period problem with integer trading constraints, minimizing the risk given a probabilistic constraint on the returns (with no transaction costs), and finding exact solutions via a branch-and-bound method for problems with up to 200 assets. They considered four different methods, of which one was able to solve the largest problems to optimality in 83\% of the cases, but the other three failed for all problems (the average run time for the largest problems was 4800 seconds). They found that solving the integer problem was harder than solving a continuous problem of the same size with cardinality constraints or minimum buy-in thresholds. 

G{\^a}rleanu and Pedersen \cite{garleanu2013dynamic} solved a continuous multi-period problem via dynamic programming, deriving a closed-form solution when the covariance matrix is positive definite, 
thereby offering insight on the properties of the solutions to the multi-period problem.
A multi-period trade execution problem was treated analytically by Almgren and Chriss \cite{almgren2001optimal}, motivating our inclusion of both temporary and permanent price-impact terms.

The connection between spin glasses and Markowitz portfolio optimization was shown by Galluccio \textit{et al.} \cite{galluccio1998rational}. The discrete multi-period problem was suggested by L\'{o}pez de Prado \cite{lopez2015generalized} as being amenable to solving using a quantum annealer. 
The contribution of this paper is to investigate the implementation and solution of a similar discrete multi-period problem on the D-Wave quantum annealer. 

\subsection{Integer formulation}
\label{sec:integer_formulation}

The portfolio optimization problem described above may be written as a quadratic integer optimization problem. We seek to maximize returns, taking into account the risk and transaction costs, including temporary and permanent market impact (the symbols are defined in Appendix~\ref{sec:symbol_definition}),

% VERSION ALIGNED FOR DOUBLE COLUMNS:
\begin{dmath}
\label{objective_w}
	w = \text{argmax}_w \sum_{t=1}^T 
	\left( \mu_{t}^T w_t - \frac{\gamma}2 w_t^T \Sigma_t w_t - \Delta w_{t}^T \Lambda_t \Delta w_{t}  
   	+  \Delta w_{t}^T \Lambda_t^{'} w_{t}  \right),
\end{dmath}

%
%\begin{align}
%\label{objective_w}
%w = \mbox{argmax}_w \sum_{t=1}^T &\left\{ \mu_{t}^T w_t - \frac{\gamma}2 w_t^T \Sigma_t w_t - \Delta w_{t}^T \Lambda_t \Delta w_{t}
% -  \Delta w_{t}^T \Lambda_t^{'} \Delta w_{t}  \right\},
%\end{align}
%
\noindent subject to the constraints that the sum of holdings at each time step be equal to $K$, 

\begin{eqnarray}
\label{constraint}
\forall t: \sum_{n=1}^{N} w_{nt} = K,
\end{eqnarray}
and that the maximum allowed holdings of each asset be $K'$,
\begin{equation}
\forall t, \forall n: w_{nt} \leq K'.
\end{equation}
The first term in Eq.~\ref{objective_w} is the sum of the returns at each time step, which is given by the forecast returns $\mu$ times the holdings $w$.
The second term is the risk, in which the forecast covariance tensor is given by $\Sigma$, and $\gamma$ is the risk aversion. The third and fourth terms encapsulate transaction costs.
Specifically, the third term includes any fixed or relative direct transaction costs, as well as the temporary market impact, while the fourth term captures any permanent market impact caused by trading activity. \hl{The transaction cost term is square in the change in the holdings, so it penalizes changes in the holdings if the corresponding entry in $\Lambda_t$ is positive \cite{garleanu2013dynamic}. The permanent market impact term allows for the fact that increasing a large holding requires executing a large buy order, which increases the price, and hence the returns \cite{almgren2001optimal}.}
%%%%%%%%%%%%%%%%%%%%%%%%%%%%%%%%%%%%%%%
\subsection{Extensions}
\label{sec:extensions}

A straightforward extension can be made to solve optimal trade execution problems. For example, in order to solve a problem in which the asset manager has $K$ units invested and would like to liquidate them over $T$ time steps, the constraint in Eq.~\ref{constraint} would change to
\begin{eqnarray}
\forall t: \sum_{n=1}^{N} w_{nt} \leq K
\end{eqnarray}
and the sum in the transaction cost and permanent impact terms would extend to time step $T+1$ with $w_{T+1} = 0$ (the zero vector).

The risk term in Eq.~\ref{objective_w} requires the estimation of a covariance matrix for each time step. There are cases in which this is problematic: for example, if not enough data exists for a good estimate or if some of the assets were not traded due to low liquidity. An alternative and more direct way to quantify risk is via the variance of the returns stream of the proposed trajectory. This avoids the issues with the estimation of covariance matrices. An additional advantage of this method is that it does not assume a normal distribution of returns. The disadvantage is that if the number of time steps $T$ is small, the estimate of the true variance of the proposed trajectory will be poor. A high variance of returns is penalized regardless of whether it occurs due to positive or negative returns.

The variance is quadratic in the returns, so it is a suitable term to include in a quadratic integer formulation. We use the identity $\mbox{Var}(r) = \langle r^2 \rangle - \langle r \rangle ^2$, and note that the returns stream is given by the vector $r[w] =\mbox{diag}(\mu^T w)$. We find the alternative risk term
\begin{dmath}
	\text{risk}[w] = \frac{\gamma}{T} \sum_{t=1}^T
	\left[ \left(\mu_{t}^T  w_{t}\right)^2  - \frac{1}{T}  \sum_{t'=1}^T \left( \mu_{t}^Tw_{t} \right) \left( \mu_{t'}^T  w_{t'} \right) \right].
\end{dmath}
%

%%%%%%%%%%%%%%%%%%%%%%%%%%%%%%%%%%%%%%%
%%%%%%%%%%%%%%%%%%%%%%%%%%%%%%%%%%%%%%%
\section{Solution using a quantum annealer}

%%%%%%%%%%%%%%%%%%%%%%%%%%%%%%%%%%%%%%%
\subsection{Quantum annealing}
\label{sec:quantum_annealing}

Quantum annealing is a process which can be used to find the optimal solution to optimization problems, if these problems can be encoded as a Hamiltonian \cite{finnila1994quantum, kadowaki1998quantum}. To this end, the quantum system is first prepared such that it represents a trivial problem, and is in the ground state of that problem, which is an equally weighted superposition of all possible states. The system is then transformed continuously to the point that it represents the optimization problem that we want to solve. If this process is done slowly enough, the adiabatic theorem guarantees that the system will remain in the ground state, as long as external disturbances are absent. The state of the system is then read, and in the ideal case it would correspond to the optimal solution of the optimization problem we wish to solve \cite{farhi2001quantum}. This process is referred to as ``adiabatic quantum computation''. In a real device, external interference always exists to some degree, so the result is probabilistic, and annealing the same problem multiple times increases the probability of finding the optimum. Therefore, quantum annealers are effectively heuristic solvers.

It has been argued that quantum annealing has an advantage over classical optimizers due to quantum tunnelling. Quantum tunnelling allows an optimizer to more easily search the solution space of the optimization problem, thereby having a higher probability of finding the optimal solution.
This might provide a speed improvement over classical optimizers, at least for certain problem classes \cite{ray1989sherrington, finnila1994quantum, kadowaki1998quantum, santoro2002theory, battaglia2005optimization}. 

D-Wave Systems has developed a scalable quantum annealer. Mathematically, this is a device which minimizes unconstrained binary quadratic functions,
\begin{align}
& \min x^TQx \\
& s.t . \, x \in \{0,1\}^N, \nonumber 
\end{align} 
where $Q \in \mathbb{R}^{N\times N}$ \cite{harris2010experimental, johnson2011quantum, boixo2013experimental}. In order to keep external disturbances to a minimum, D-Wave Systems' quantum annealer is cooled to 15 mK (about 180 times colder than interstellar space), is shielded from RF signals due to its being housed inside a metal enclosure, is shielded from external magnetic fields larger than 1 nT (about 50,000 times less than Earth's magnetic field), and operates in a high-vacuum environment in which the pressure is 10 billion times lower than atmospheric pressure \cite{DWave2XQuantumComputer}. 

There is strong evidence that the D-Wave machine is indeed quantum \cite{lanting2014entanglement, boixo2014computational}. Recently, there has been significant interest in benchmarking the D-Wave machines using different metrics, and often against classical solvers \cite{mcgeoch2013experimental, boixo2014evidence, katzgraber2014glassy, hen2015probing, king2015performance, king2015benchmarking}. There is an ongoing debate on how to define quantum speedup, and on which problems a noisy quantum annealer would be expected to show such a speedup \cite{ronnow2014defining, martin2015unraveling, crosson2016simulated}. \hl{Recently, Denchev \textit{et al.} claimed a $10^8$ speedup over simulated annealing when solving a specially constructed class of problems on a single-core machine \cite{denchev2015computational}.} It is still an open question whether D-Wave Systems' quantum annealer shows a quantum speedup. We expect new results to shed light on this in the near future.

\begin{figure}[b]
\caption{An example hardware graph, showing the connectivity of the qubits for a Chimera graph with $s=4$ unit cells in each row/column, giving a total of $q=128$ qubits.}
\centering
\includegraphics[scale=.55]{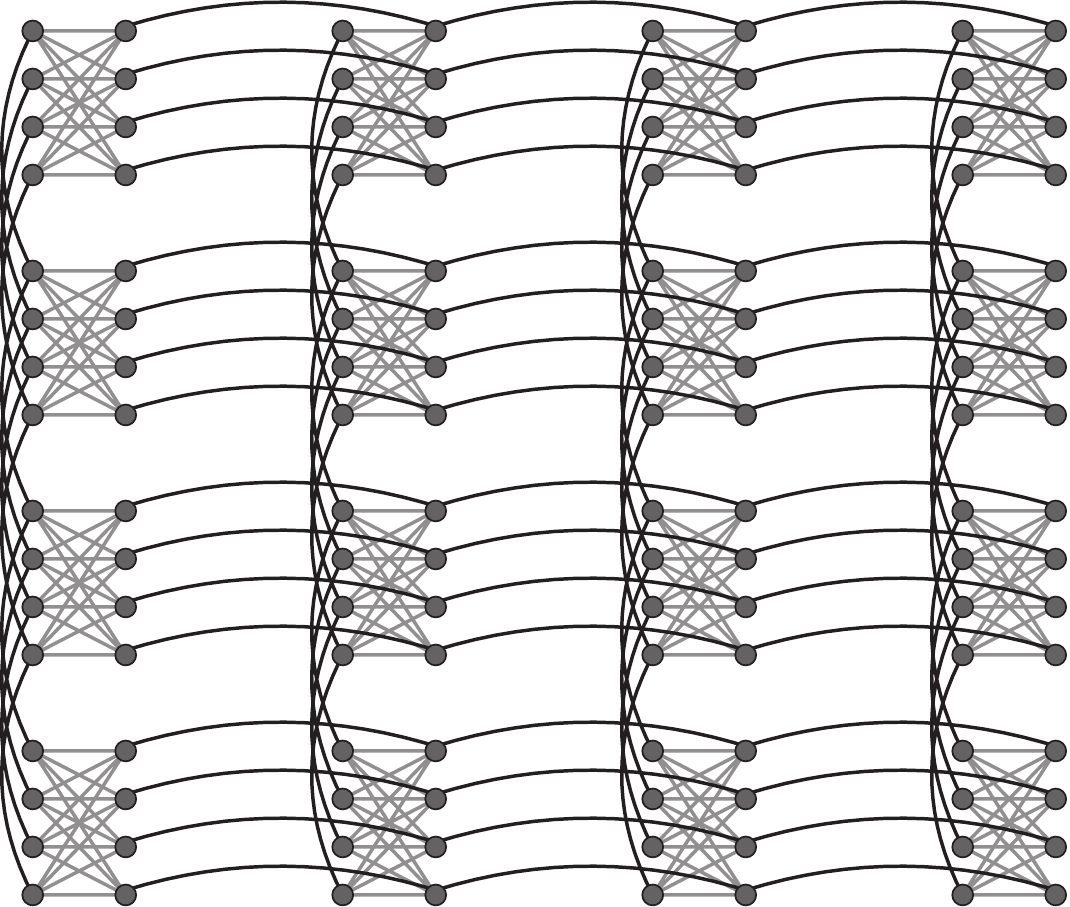}
\label{fig:chimera}
\end{figure}

The connectivity of the qubits in D-Wave's quantum annealer is currently described by a square Chimera graph \cite{bunyk2014architectural}. This hardware graph is composed of a lattice of bipartite unit cells containing eight qubits. Qubits in adjacent unit cells are connected if they are in the same position in the unit cell (see Fig.~\ref{fig:chimera} for an example).  

%
% NOTE: figure created using Mohammad Vazifeh's code located at: git/PY_Algorithmic.repo/1QBit.MachineLearning/hopfieldNetwork/Chimera.py

If we label the number of unit cells along an edge as $s$, then the total number of qubits is $q=8s^2$. The hardware graph is sparse and in general does not match the problem graph, which is defined by the adjacency matrix of the problem matrix $Q$. In order to solve problems that are denser than the hardware graph, we identify multiple physical qubits with a single logical qubit (a problem known as ``minor embedding'' \cite{choi2008minor, choi2011minor}), at the cost of using many more physical qubits. For square Chimera hardware graphs, the size $V$ of the largest fully dense problem that can be embedded on a chip with $q$ qubits is $V=\sqrt{2q}+1=4s+1$, assuming no faulty qubits or couplers. For example, the latest chip is the \mbox{D-Wave} 2X\footnote{\hl{as of April 2016}}, which has $s=12$ unit cells along each side, giving $q=1152$ qubits, for which we get $V=49$. \hl{Lower-density problems of significantly larger size can be embedded.  For example, on one of the annealers used in this study, which has 1100 qubits, problems with $V_b\simeq140$ and a density of $\simeq0.1$ were able to be embedded.}

%%%%%%%%%%%%%%%%%%%%%%%%%%%%%%%%%%%%%%%
\subsection{From integer to binary}
\label{sec:integer2binary}

To solve this problem using the D-Wave quantum annealer, the integer variables of Eq.~\ref{objective_w} must be recast as binary variables, and the constraints must be incorporated into the objective function. We have investigated four different encodings: binary, unary, sequential, and partitioning.  The first three can be described by writing the integer holdings as a linear function of the binary variables,
\begin{equation}
\label{general_encoding_eq0}
w_{nt}[x] = \sum_{d=1}^{D} f(d) x_{dnt},
\end{equation}
where $x_{dnt} \in \{0,1\}$ and the encoding function $f(d)$ and the bit depth $D$ for each encoding are given in Table~\ref{table:encoding_functions}.

\begin{table}[t]
\caption{Encodings: $f(d)$ is the encoding function and $D$ is the bit depth.}
\begin{center}
  \begin{tabular}{c l l}
    \hline
    Encoding & $f(d)$ & $D$ \\ \hline \hline
    Binary        & $2^{d-1}$ & $\mbox{log}_2(K'+1)$                     \\ \hline
    Unary        & $1$           & $K'$                                                 \\ \hline
    Sequential & $d$          & $\left( \sqrt{1+8K'}-1 \right)/2$ \\ \hline
  \end{tabular}
\end{center}
\label{table:encoding_functions}
\end{table}

The fourth encoding involves finding all partitions of $K$ into $N$ assets with $K'$ or less units in each asset, and assigning a binary variable to each encoding at each time step.

%\newgeometry{margin=1cm} % modify this if you need even more space
%\begin{landscape}
\begin{table*}[t]
%\begin{center}
\caption{Comparison of the four encodings described in Section~\ref{sec:integer2binary}. The column ``Variables'' refers to the number of binary variables required to represent a particular problem. The column ``Largest integer'' refers to a worst-case estimate of the largest integer that could be represented based on the limitation imposed by the noise level $\epsilon$ and the ratio between the largest and smallest problem coefficients $\delta$ and $n \equiv 1 / \sqrt{\epsilon \delta}$.}
\begin{tabular}{llll}\hline
Encoding & Variables & Largest integer & Notes \\ \hline
Binary & $TN\log_2(K'+1)$ &  $\left\lfloor 2n \right\rfloor - 1$ & \parbox[c]{6.35cm}{Most efficient in number of variables; allows representing of the second-lowest integer.} \\ \hline
Unary &  $TNK'$ & No limit & \parbox[c]{6.35cm}{Biases the quantum annealer due to differing redundancy of code words for each value; encoding coefficients are even, giving no dependence on noise, so it allows representing of the largest integer.} \\ \hline
Sequential & $\frac12 TN \left( \sqrt{1+8K'}-1 \right)$ & $\frac12 \left\lfloor n \right\rfloor \left( \left\lfloor n \right\rfloor + 1\right)$ & \parbox[c]{6.35cm}{Biases the quantum annealer (but less than unary encoding); second-most-efficient in number of variables; allows representing of the second-largest integer.} \\ \hline
Partition & $\leq T{K+N-1 \choose N-1}$ &  $\left\lfloor n \right\rfloor$ &  \parbox[c]{6.35cm}{Can incorporate complicated constraints easily; least efficient in number of variables; only applicable for problems in which groups of variables are required to sum to a constant; allows representing the lowest integer.} \\ \hline
\end{tabular}
\label{encoding_comparison}
%\end{center}
\end{table*}
%\end{landscape}
%\restoregeometry

\begin{table*}[t]
\caption{Dependence of the number of binary variables required on the number of units $K$, number of assets $N$, and the number of time steps $T$ for some example values (here we assumed $K'=K/3$). The number of variables is given for the three linear encodings: binary $V_b$, unary $V_u$, and sequential $V_s$. } 
\begin{center}
  \begin{tabular}{rrrrrrr}
\hline
   $N$ &   $T$ &  $K$ &   $K'$ &   $V_b$ &   $V_u$ &  $V_s$ \\
\hline
   5 &   5 &  15 &    5 &    75 &   125 &    75 \\
  10 &  10 &  15 &    5 &   300 &   500 &   300 \\
  10 &  15 &  15 &    5 &   450 &   750 &   450 \\
  20 &  10 &  15 &    5 &   600 &  1000 &   600 \\
  50 &   5 &  15 &    5 &   750 &  1250 &   750 \\
  20 &  15 &  15 &    5 &   900 &  1500 &   900 \\
  50 &  10 &  15 &    5 &  1500 &  2500 &  1500 \\
  50 &  15 &  15 &    5 &  2250 &  3750 &  2250 \\
\hline
\end{tabular}
\end{center}
%The density $d$ of the given portfolio optimization problem is given by Eq.~\ref{density_po}.}
\label{num_variables_table}
\end{table*}

We summarize the properties of each of the four encodings described above in Table~\ref{encoding_comparison}.
Which encoding is preferred will depend on the problem being solved and the quantum annealer being used. Table~\ref{num_variables_table} presents the number of variables required for some example multi-period portfolio optimization problems for each of these encodings. 

For binary, unary, and sequential encodings, there is a trade-off between the efficiency of the encoding---the number of binary variables needed to represent a given problem---and the largest integer that can be represented. The reason is that for an encoding to be efficient it will typically introduce large coefficients, limiting the largest integer representable (due to the noise level in the quantum annealer). For example, the binary encoding is the most efficient of the three (that is, it requires the fewest binary variables); however, it is the most sensitive to noise, and hence can represent the smallest integer of the three, given some number of qubits. Conversely, the unary encoding is the worst of the three in efficiency, but can represent the largest integer. An additional consideration is that some encodings introduce a redundancy that biases the quantum annealer towards certain solutions. Briefly, each integer can be encoded in multiple ways, the number of which is (in general) different for each integer. In this scenario, the quantum annealer is biased towards integers that have a high redundancy. 

The partition encoding is different in that it requires an exponential number of variables; however, it allows a straightforward formulation of complicated constraints, like cardinality, by excluding partitions that break the constraints, which also lowers the number of variables required. For the other three encodings, constraints can be modelled through the encoding (for example, a minimum or maximum holdings constraint), or through linear or quadratic penalty functions. We note that the actual number of physical qubits required could be much larger than the number of variables indicated in Table~\ref{num_variables_table} due to the embedding (see Section~\ref{sec:quantum_annealing}).

\begin{table*}[t]
\caption{Results using D-Wave's \hl{512-qubit} quantum annealer (with $200$ instances per problem): $N$ is the number of assets, $T$ is the number of time steps, $K$ is the number of units to be allocated at each time step and the maximum allowed holding (with $K'=K$), ``encoding'' refers to the method of encoding the integer problem into binary variables, ``vars'' is the number of binary variables required to encode the given problem, ``density'' is the density of the quadratic couplers, ``qubits'' is the number of physical qubits that were used, ``chain'' is the maximum number of physical qubits identified with a single binary variable, and $S(\alpha)$ refers to the success rate given a perturbation magnitude $\alpha\%$ (explained in the text).}
\begin{center}
  \begin{tabular}{r r r l r r r r r r r}
    \hline
 $N$ & $T$ & $K$ & encoding & vars & density & qubits & chain &  $S(0)$ &  $S(1)$ &  $S(2)$ \\ \hline
   2 &   3 &   3 & binary     &     12 &      0.52 &       31 &       3 &   100.00 &   100.00 &   100.00 \\
   2 &   2 &   3 & unary      &     12 &      0.73 &       45 &       4 &    97.00 &    99.50 &   100.00 \\
   2 &   4 &   3 & binary     &     16 &      0.40 &       52 &       4 &    96.00 &   100.00 &   100.00 \\
   2 &   3 &   3 & unary      &     18 &      0.53 &       76 &       5 &    94.50 &    99.50 &   100.00 \\
   2 &   2 &   7 & binary     &     12 &      0.73 &       38 &       4 &    90.50 &   100.00 &   100.00 \\
   2 &   5 &   3 & binary     &     20 &      0.33 &       63 &       4 &    89.00 &   100.00 &   100.00 \\
   2 &   6 &   3 & binary     &     24 &      0.28 &       74 &       4 &    50.00 &    97.50 &    99.50 \\
   3 &   2 &   3 & unary      &     18 &      0.65 &       91 &       6 &    38.50 &    72.50 &    91.50 \\
   3 &   3 &   3 & binary     &     18 &      0.45 &       84 &       5 &    35.50 &    66.50 &    82.50 \\
   3 &   4 &   3 & binary     &     24 &      0.35 &      106 &       6 &     9.50 &    50.50 &    84.50 \\ \hline
\end{tabular}
\end{center}
\label{table:results_system13_raw}
\end{table*}

In many cases, the maximum holdings $K'$, which is also the largest integer to be represented, will not be exactly encodable using the binary and sequential encodings. For example, using a binary encoding one can encode the values $0$ to $7$ using three bits, and $0$ to $15$ using four bits, but integers with a maximum value between $7$ and $15$ are not exactly encodable. In order to avoid encoding infeasible holdings, these can be penalized by an appropriate penalty function. However, this penalty function will typically be a high-order polynomial and require many auxiliary binary variables in order to reduce it to a quadratic form. Instead, we propose to modify the encoding by adding bits with the specific values needed. For example, $\{1, 1, 2, 2, 4\}$ represents a modified binary encoding for the values $0$ to $10$.

The constraints of Eq.~\ref{constraint} can be incorporated into the objective function by rearranging the equations, squaring them, and summing the result, obtaining the penalty term
\begin{equation}
\mbox{penalty}[w] = -M\sum_{t=1}^{T} \left(K - \sum_{n=1}^{N} w_{nt} \right)^2,
\end{equation}
where $M>0$ is the strength of the penalty. In theory, $M$ can be chosen large enough such that all feasible solutions have a higher value than all infeasible solutions. In practice, having an overly large $M$ leads to problems due to the noise in the system (see Section~\ref{sec:discussion}), so we choose it empirically by trial and error. In the future, it might be possible to include equality constraints of this form via a change in the quantum annealing process, allowing us to drop this term \cite{hen2015quantum}.

We note that an alternative approach, involving the tiling of the integer search space with binary hypercubes, was investigated by \cite{ronagh2013integer}; however, it requires an exponential number of calls to the quantum annealer. In addition, a 

%%%%%%%%%%%%%%%%%%%%%%%%%%%%%%%%%%%%%%%
\subsection{Numerical Results}
\label{sec:results}

\begin{table*}[t]
\caption{Results \hl{using D-Wave's 512-qubit quantum annealer}, with custom-tuned parameters and software (with $200$ instances per problem): an improved embedding solver, fine-tuned identification coupling strengths, and averaging over $5$ random gauges ($200$ reads per gauge, giving a total of $1000$ reads per call). Columns are as in Table~\ref{table:results_system13_raw}, and the hardware used was the same as in that table.}
\begin{center}
  \begin{tabular}{r r r l r r r r r r r}
    \hline
 $N$ & $T$ & $K$ & encoding & vars & density & qubits & chain &  $S(0)$ &  $S(1)$ &  $S(2)$ \\ \hline
   2 &   3 &   3 & binary     &     12 &      0.52 &       31 &       3 &   100.00 &   100.00 &   100.00 \\
   2 &   4 &   3 & binary     &     16 &      0.40 &       52 &       4 &    99.50 &   100.00 &   100.00 \\
   3 &   2 &   3 & unary      &     18 &      0.65 &       91 &       6 &    99.00 &   100.00 &   100.00 \\
   2 &   3 &   3 & unary      &     18 &      0.53 &       76 &       5 &    98.50 &    99.50 &   100.00 \\
   2 &   5 &   3 & binary     &     20 &      0.33 &       63 &       4 &    96.00 &   100.00 &   100.00 \\
   2 &   6 &   3 & binary     &     24 &      0.28 &       74 &       4 &    80.00 &   100.00 &   100.00 \\
   2 &   4 &   3 & unary      &     24 &      0.41 &      104 &       6 &    70.50 &    96.50 &    99.50 \\
   3 &   4 &   3 & binary     &     24 &      0.35 &      106 &       6 &    44.00 &    92.50 &    99.00 \\
   3 &   3 &   3 & binary     &     18 &      0.45 &       84 &       5 &    65.50 &    98.00 &   100.00 \\
   3 &   6 &   3 & binary     &     36 &      0.24 &      196 &       7 &     0.50 &    74.50 &    99.00 \\
   4 &   4 &   3 & binary     &     32 &      0.32 &      214 &       8 &     1.50 &    13.50 &    60.50 \\
   4 &   5 &   3 & binary     &     40 &      0.26 &      281 &      10 &     0.00 &     3.00 &    24.00 \\ \hline
\end{tabular}
\end{center}
\label{table:results_improved_software}
\end{table*}
\begin{table*}[t]
\caption{Results using \hl{D-Wave's 1152-qubit quantum annealer (with $200$ instances per problem), with lower noise and higher yield, and with custom-tuned parameters and software (as in Table~\ref{table:results_improved_software})}. Columns are as in Table~\ref{table:results_system13_raw}.}
\begin{center}
\hlc{
  \begin{tabular}{r r r l r r r r r r r}
    \hline
 $N$ & $T$ & $K$ & encoding & vars & density & qubits & chain &  $S(0)$ &  $S(1)$ &  $S(2)$ \\ \hline
   3 &   2 &   3 & unary      & 18     &      0.65 &       86 &       5 &    99.50 &   100.00 &   100.00 \\
   3 &   3 &   3 & binary     & 18     &      0.45 &       61 &       4 &    83.50 &    99.00 &   100.00 \\
   3 &   3 &   3 & unary      & 27     &      0.46 &      146 &       7 &    81.50 &    92.50 &    97.00 \\
   3 &   4 &   3 & binary     & 24     &      0.35 &       87 &       5 &    75.50 &   100.00 &   100.00 \\
   3 &   4 &   3 & unary      & 36     &      0.36 &      209 &       8 &    40.50 &    52.50 &    60.50 \\
   4 &   4 &   3 & binary     & 32     &      0.32 &      157 &       6 &    27.00 &    46.50 &    64.50 \\
   3 &   6 &   3 & binary     & 36     &      0.24 &      143 &       5 &    15.50 &    59.00 &    78.50 \\
   4 &   5 &   3 & binary     & 40     &      0.26 &      210 &       7 &     4.00 &    34.50 &    64.00 \\
   5 &   5 &   3 & binary     & 50     &      0.25 &      321 &       8 &     4.00 &    13.00 &    27.00 \\
   6 &   5 &   3 & binary     & 60     &      0.24 &      492 &      10 &     2.00 &    20.50 &    49.50 \\
   6 &   6 &   3 & binary     & 72     &      0.20 &      584 &      11 &     0.00 &     6.50 &    21.00 \\   \hline
\end{tabular}}
\end{center}
\label{table:results_improved_chip}
\end{table*}

The results for a range of portfolio optimization problems are presented in Table~\ref{table:results_system13_raw}.
\hl{For each problem $200$ random instances were generated. Each instance was solved using one query to the quantum annealer, with $1000$ reads per query (which involved either 1 call or 5 calls if averaging over gauges)\footnote{\hl{We have observed that the success rate rises when increasing the number of reads, for a fixed problem size. If the number of reads is fixed, the success rate is expected to decrease as problem size increases, as seen in the results.}}. All results were obtained from chips with a hardware graph with either 512 or 1152 qubits.  (The number of active qubits was a little smaller.)} For validation purposes, each instance was also solved using an exhaustive integer solver to find the optimal solution. For the larger problems, a heuristic solver was run a large number of times in order to find the optimal solution with high confidence.

As a solution metric, we used the percentage of instances for each problem for which the quantum annealer's result fell within perturbation magnitude $\alpha\%$ of the optimal solution, denoted by $S(\alpha)$. This metric was evaluated by perturbing each instance at least 100 times, by adding Gaussian noise with standard deviation given by $\alpha\%$ of each eigenvalue of the problem matrix $Q$. Each perturbed problem was solved by an exhaustive solver, and the optimal solutions were collected. If the quantum annealer's result for that instance fell within the range of optimal values collected, then it was deemed successful (within a margin of error). This procedure was repeated for each random problem instance, giving a total success rate for that problem. For the case $\alpha=0$, this reduces to defining success as the finding of the optimal solution.

\hl{The chosen approach relaxes the success metric in a problem-instance-specific way.  An alternative would be to define success as finding a solution within $\epsilon$ of the optimum. However this alternative success metric has the disadvantage that $\epsilon$ could be small or large compared to the energy scale of a particular problem instance, and} \hl{so the metric can be misleading} \hl{for problems of the type being solved here.}

\hl{Although the variance of the success rate would also be interesting to observe, the number of experimental runs required for this metric to be statistically significant were not able to be performed due to a limited availability of machine time.}

In order to investigate the quality of the solution on software enhancements, we replaced the ``embedding solver'' supplied with the D-Wave quantum annealer with a proprietary embedding solver developed by 1QBit, tuned the identification coupling strength, and combined results from calls with multiple random gauges. A gauge transformation is accomplished by assigning $+1$ or $-1$ to each qubit, and flipping the sign of the coefficients in the problem matrix accordingly, such that the optimization problem remains unchanged. We found a large improvement for all problems. Results with these improvements are presented in Table~\ref{table:results_improved_software}.

To investigate the dependence of the success rate on the noise level and qubit yield of the quantum annealer, several problems were solved on two different quantum annealing chips. Results for the second chip, \hl{the D-Wave 2X, which has 1152 qubits, }a lower noise level and fewer inactive qubits and couplers, are presented in Table~\ref{table:results_improved_chip}. We found an increase in all success rates, as well as the ability to solve larger problems. 

These investigations highlight the importance of having an in-depth understanding of D-Wave's quantum annealer in order to be able to achieve the best results possible.

%%%%%%%%%%%%%%%%%%%%%%%%%%%%%%%%%%%%%%%
\subsection{Discussion}
\label{sec:discussion}

The success rate and the ability to solve larger problems are affected by certain hardware parameters of the quantum annealer. First, there is a level of intrinsic noise which manifests as a misspecification error---the coefficients of the problem that the quantum annealer actually solves differ from the problem coefficients by up to $\epsilon$. \footnote{\hl{The intrinsic noise for the current generation of chip is estimated to be around 2\%--4\% of the full scale.}} For future chip generations the expectation is that $\epsilon$ will decrease, and hence the success rate will increase by virtue of the quantum annealer solving a problem that is closer to the problem passed to it.  In addition, the problem coefficients on the chip have a defined coefficient range, and if the specified problem has coefficients outside this range, the entire problem is scaled down. This can result in coefficients becoming smaller than $\epsilon$, affecting the success rate. These factors are especially relevant for high-precision problems such as the multi-period portfolio optimization problem solved here.

The quantum annealer has a hardware graph that is currently very sparse and in general does not match the problem graph. In order to solve problems that are denser than the hardware graph, multiple physical qubits are identified with a single binary variable, at the cost of using more physical qubits. In order to force the identified qubits to all have the same value, a strong coupling is needed. If the required coupling is outside of the range of the couplings in the problem, the result will be an additional scaling, possibly reducing additional coefficients to less than $\epsilon$, again impacting the success rate.  Generally, the denser the hardware graph is, the fewer identifications are needed, and the weaker the couplings are required to be to identify the qubits. 

To solve larger problems, the number of qubits must be greater. More qubits would also allow the use of an integer encoding scheme that is less sensitive to noise levels (such as unary encoding versus binary encoding). The fabrication process is not perfect, resulting in inactive qubits and couplers. The more inactive qubits and couplers there are on a chip, the lower the effective density and the higher the number of identifications required, which typically reduces the success rate. 

In addition, custom tuning, through software, can be used to enhance the results. In particular, when the problem to be solved has a graph that differs from the hardware graph, a mapping, referred to as an ``embedding'', must be found from the problem graph to the hardware graph. The development of sophisticated ways to find better embeddings (for example, with fewer identified qubits) would be expected to increase the success rate---often the structure of the problem can be exploited in order to find better embeddings. In addition, when an embedding is used, there are different ways in which the couplings of the identified qubits should be set, controlled by the ``embedding solver''. For example, the strength of the couplings could be fine-tuned further to give higher success rates, or tuned separately for each set of identified qubits \cite{perdomo2015performance}.

 \hl{The issue of which embedding properties are most desirable is still under active research.  Here the pi-elite metric has been used to select the best scaling of the problem versus the strength of the qubit identification chains, and to choose the best embedding amongst the highest-ranked embeddings \cite{perdomo2015performance}.  The pi-elite score is determined by comparing the mean energy of the best (``elite'') states (for example, the lowest 2\%) found by using each scale/embedding.  The embeddings were ranked using a scheme that combines equal weights based on the shortest of the longest chains, the total number of qubits, and the variance of the lengths of the chains, after which the highest-ranked embeddings were compared using their pi-elite scores.}
 
In addition it has been observed that there is a gauge transformation under which the problem and solution remain invariant, but the success rates vary strongly (due to imperfections in the annealing chip). Combining results from multiple calls to the solver with random gauges, as we did, could provide a large improvement \cite{perdomo2015performance}. Software error correction, such as majority voting (which we employed) when the identified physical qubits do not agree, as well as calibration, could also lead to improved solutions \cite{mishra2015performance, pudenz2014error, pudenz2015quantum, vinci2015quantum, perdomo2015determination, pastawski2015error, lechner2015quantum}. We also note that it may be possible to use the quantum annealer to find good local minima, which could then be used to speed up deterministic or heuristic classical solvers \cite{katzgraber2015seeking,zaribafiyan2015quamip}.

\hl{
Although the core contributions of this paper are the formulation of the general multi-period optimization strategy and discussion of the issues involved in solving that problem using available quantum annealing hardware, a brief comment regarding the time taken to calculate a solution is warranted.  In general, benchmarking of the time to solution of a D-Wave quantum annealer against classical hardware is an area of ongoing and active research.  For the small-scale problems solved in this study, the time to solution, on both classical hardware and using the quantum annealer, is comparable.  The recent results by Denchev \textit{et al.} \cite{denchev2015computational}, which show that for a specific class of problems the D-Wave machine has the potential to provide speedup over computations on classical hardware, are encouraging.
However, only after quantum speedup has been demonstrated for general problems, and specifically those requiring a high precision of couplings, is it expected that a quantum speedup for the optimal trading trajectory problem will be observed.
}

%%%%%%%%%%%%%%%%%%%%%%%%%%%%%%%%%%%%%%%
%%%%%%%%%%%%%%%%%%%%%%%%%%%%%%%%%%%%%%%
\section{Conclusions}
\label{sec:conclusions}

In this limited experiment we have demonstrated the potential of D-Wave's quantum annealer to achieve high success rates when solving an important and difficult multi-period portfolio optimization problem. We have also shown that it is possible to achieve a considerable improvement in success rates by fine-tuning the operation of the quantum annealer.

Although the current size of problems that can be solved is small, technological improvements in future generations of quantum annealers are expected to provide the ability to solve larger problems, and at higher success rates.

\begin{table*}[t]
\caption{Definition of symbols}
\begin{center}
  \begin{tabular}{c l l}
    \hline
    Symbol & Type & Description \\ \hline
    $K$ & $\mathbb{N}_1$ & Number of units to be allocated at each time step \\ 
     $K'$ & $\mathbb{N}_1$ & Largest allowed holding for any asset \\ 
    $N$ & $\mathbb{N}_1$ & Number of assets \\ 
    $T$ & $\mathbb{N}_1$ & Number of time steps \\
    $\mu$ & $\mathbb{R}^{N \times T}$ & Forecast mean returns of each asset at each time step \\
    $\gamma$ & $\mathbb{R}$ & Risk aversion \\ 
    $\Sigma$ & $\mathbb{R}^{T \times N \times N}$ & Forecast covariance matrix for each time step \\
    $c'$ & $\mathbb{R}^{N \times T}$ & Permanent market impact coefficients for each asset at each time step \\
    $c$ & $\mathbb{R}^{N \times T}$ & Transaction cost coefficients for each asset at each time step \\
    $w_0$ & $\mathbb{N}_0^N$               & Initial holdings for each asset \\ 
    $w_{T+1}$ & $\mathbb{N}_0^N$           & Final holdings for each asset \\
    $w$     & $\mathbb{N}_0^{N\times T}$ & Holdings for each asset at each time step (the trading trajectory) \\ \hline
  \end{tabular}
\end{center}
\label{table:variable_definition}
\end{table*}

Since larger problems are expected to be intractable on classical computers, there is much interest in solving them efficiently using quantum hardware.

\appendices

\section{Definition of Symbols}
\label{sec:symbol_definition}

The symbols used above are defined in Table~\ref{table:variable_definition}. In addition, we use $w_t$ to denote the $t$-th column of the matrix $w$ (and similarly for $\mu$), and $\Sigma_t$ to denote the covariance matrix ($N \times N$) which is the $t$-th page of the tensor $\Sigma$. For convenience of notation, the temporary transaction costs $c$ are represented using the tensor $\Lambda$, where $\Lambda_{tnn'}=c_{nt}\delta_{nn'}$ (and similarly for the permanent price impact $c'$ and $\Lambda'$). The difference in holdings between two time periods is defined as $\Delta w_t \equiv w_t-w_{t-1}$.

%\section{}

\section*{Acknowledgment}

The authors would like to thank Marko Bucyk for editing a draft of this paper, and Robyn Foerster, Pooya Ronagh, Maxwell Rounds, and Arman Zaribafiyan for their useful comments. \hl{Clemens Adolphs and Dominic Marchand provided invaluable help in obtaining the numerical results, and Mohammad Vazifeh's code was used for the Chimera-graph figure.} This work was supported by 1QB Information Technologies and Mitacs.

\ifCLASSOPTIONcaptionsoff
  \newpage
\fi



%\bibliographystyle{IEEEtran}

\bibliographystyle{IEEEtran}
\bibliography{IEEEabrv,OptimalTrading_WHPCF}

%\begin{thebibliography}{1}

%\end{thebibliography}


% If no photo then use
%\begin{IEEEbiographynophoto}{Gili Rosenberg}

\begin{IEEEbiography}[{\includegraphics[width=1in,height=1.25in,clip,keepaspectratio]{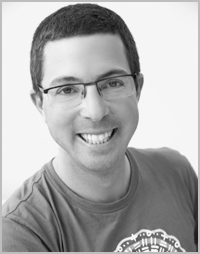}}]{Gili Rosenberg} is a Senior Researcher at 1QBit.  His research focuses on the application of quantum annealing technology to solve problems in computational finance that are traditionally formulated using machine learning techniques such as neural networks, and heuristic algorithms such as simulated annealing.   

Dr. Rosenberg holds a Ph.D. in Physics from the University of British Columbia (2012) and a B.Sc. in Physics from Ben-Gurion University (2004).
\end{IEEEbiography}

\begin{IEEEbiography}[{\includegraphics[width=1in,height=1.25in,clip,keepaspectratio]{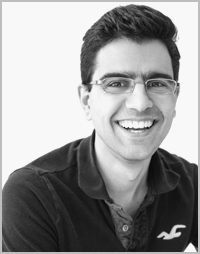}}]{Poya Haghnegahdar}
\hl{is currently completing a Ph.D. degree in Physics at the University of British Columbia where his research focuses on the measurement-based model of quantum computing, Monte Carlo simulation techniques, and percolation theory.

He was formerly the Applications and QA Lead at 1QBit and has practical experience in machine learning and data analysis techniques.}
He holds an M.Sc. in Physics from the University of British Columbia (2009) and a B.Sc. in Physics from the University of Waterloo (2005).

\end{IEEEbiography}

\begin{IEEEbiography}[{\includegraphics[width=1in,height=1.25in,clip,keepaspectratio]{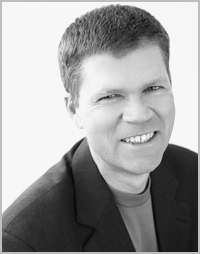}}]{Phil Goddard}
 is Head of Research at 1QBit.  His research focuses on the application of heuristic and machine learning algorithms to topics found in computational finance and traditional engineering.  Working with clients and partners spanning a range of business sectors, he leads teams focused on applying quantum annealing technology to a variety of traditional optimization problems.

With 25 years of experience in designing and developing custom software applications for modelling, simulation, data analysis, and visualization, Dr. Goddard is also President and Principal Consultant at Goddard Consulting.  He was formerly a Senior Consultant at The MathWorks, where he worked with clients in the financial services, automotive, aerospace, telecommunications, and process industries.  He has also worked for British Aerospace (Dynamics) and Woodside Offshore Petroleum.  Since 2006 he has held a position as Visiting Lecturer within the Beedie School of Business at Simon Fraser University, where he lectures on topics in numerical analysis and software development for computational finance.

Dr. Goddard holds a Ph.D. in Engineering from Cambridge University (1995), and a B.Eng. from the University of Western Australia (1991).
\end{IEEEbiography}

\begin{IEEEbiography}[{\includegraphics[width=1in,height=1.25in,clip,keepaspectratio]{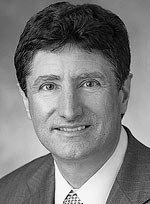}}]{Peter Carr}
\hl{currently serves as the Executive Director of the Math Finance program at NYU's Courant Institute.
Formerly a Managing Director and Global Head of Market Modeling at Morgan Stanley, Dr. Carr has approximately 20 years of experience in the financial services industry. He is also a trustee for the Museum of Mathematics in New York.}

Prior to joining the financial industry, Dr. Carr was a finance professor for 8 years at Cornell University, after obtaining his \hl{Ph.D.} from UCLA in 1989. He has over 75 publications in academic and industry-oriented journals and serves as an associate editor for 8 journals related to mathematical finance. He was selected as Quant of the Year by Risk Magazine in 2003 and Financial Engineer of the Year by IAFE and Sungard in 2010. For the last 4 years, Dr. Carr was included in Institutional Investor's Tech 50, an annual listing of the 50 most influential people in financial technology.

\end{IEEEbiography}

%\newpage

\begin{IEEEbiography}[{\includegraphics[width=1in,height=1.25in,clip,keepaspectratio]{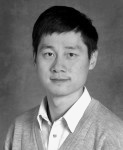}}]{Kesheng Wu}
is the Group Leader of the Scientific Data Management Group at Lawrence Berkeley National Laboratory (U.S. Department of Energy's Office of Science).  His current research focuses on indexing technology for searching large data sets, including improving bitmap index technology with compression, encoding, and binning. Dr. Wu is the key developer of FastBit bitmap indexing software, which is used in a number of applications including high-energy physics, combustion, network security, and query-driven visualization.  He has also worked on various scientific computing projects including developing the Thick-Restart Lanczos (TRLan) algorithm for solving eigenvalue problems and devising statistical tests for deterministic effects in broad band time series.

Dr. Wu holds a Ph.D. in Computer Science from the University of Minnesota, an M.Sc. in Physics from the University of Wisconsin-Milwaukee, and a B.Sc. in Physics from Nanjing University.
\end{IEEEbiography}

\begin{IEEEbiography}[{\includegraphics[width=1in,height=1.25in,clip,keepaspectratio]{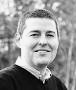}}]{Marcos L\'{o}pez de Prado}
 is Senior Managing Director at Guggenheim Partners. He is also a Research Fellow at Lawrence Berkeley National Laboratory's Computational Research Division (U.S. Department of Energy's Office of Science), where he conducts unclassified research in the mathematics of large-scale financial problems and supercomputing.
 
Previously, Dr. L\'{o}pez de Prado was Head of Quantitative Trading and Research at Hess Energy Trading Company (the trading arm of Hess Corporation, a Fortune 100 company) and Head of Global Quantitative Research at Tudor Investment Corporation. In addition to his 17 years of trading and investment management experience at some of the largest corporations, he has received several academic appointments, including Postdoctoral Research Fellow of RCC at Harvard University and Visiting Scholar at Cornell University.
 
Dr. L\'{o}pez de Prado holds a Ph.D. in Financial Economics (2003), and a second Ph.D. in Mathematical Finance (2011) from Complutense University. He is a recipient of the National Award for Excellence in Academic Performance by the Government of Spain (National Valedictorian, 1998) among other awards, and was admitted into American Mensa with a perfect test score.

He serves on the Editorial Board of the Journal of Portfolio Management (IIJ), the Journal of Investment Strategies (Risk) and the Big Data and Innovative Financial Technologies Research Series (SSRN). He has collaborated with many leading academics, resulting in some of the most read papers in finance (SSRN), four international patent applications on high frequency trading, three textbooks, and numerous publications in top mathematical finance journals.
\end{IEEEbiography}


%\vfill


\end{document}